\documentclass[fleqn,usenatbib]{mnras}

\usepackage{newtxtext,newtxmath}

\usepackage[T1]{fontenc}
\usepackage{ae,aecompl}

\usepackage{graphicx}	
\usepackage{amsmath}	
\usepackage[normalem]{ulem}
\useunder{\uline}{\ul}{}
\usepackage{placeins}
\usepackage{rotating}
\usepackage{float}
\usepackage{threeparttable}

\title[Spectroscopic confirmation of VVV CL001 with MUSE]{Spectroscopic analysis of VVV CL001 cluster with MUSE\thanks{Based on observations taken within the Program ID 60.A-9182(A)}}

\author[J. Olivares Carvajal et al.]{
J. Olivares Carvajal,$^{1,2}$\thanks{E-mail: jrolivares@uc.cl}
M. Zoccali,$^{1,2}$
A. Rojas-Arriagada,$^{1,2}$
\newauthor
R. Contreras Ramos,$^{1,2}$
F. Gran,$^{1,2}$
E. Valenti$^{3,4}$
and J. H. Minniti$^{1,2,5}$
\\
$^{1}$Instituto de Astrof\'isica, Pontificia Universidad Cat\'olica de Chile, Av. Vicu\~na Mackenna 4860, 782-0436 Macul, Santiago, Chile\\
$^{2}$Millennium Institute of Astrophysics, Av. Vicu\~na Mackenna 4860, 82-0436 Macul, Santiago, Chile \\
$^{3}$European Southern Observatory, Karl Schwarzschild-Strabe 2, D-85748 Garching bei Munchen, Germany \\
$^{4}$Excellence Cluster ORIGINS, Boltzmann-Strasse 2, D-85748 Garching Bei Munchen, Germany\\
$^{5}$Nicolaus Copernicus Astronomical Center, Polish Academy of Sciences, Bartycka 18, 00-716 Warsaw, Poland
}

\date{Accepted XXX. Received YYY; in original form ZZZ}

\pubyear{2022}

\begin{document}
\label{firstpage}
\pagerange{\pageref{firstpage}--\pageref{lastpage}}
\maketitle

\begin{abstract}
Like most spiral galaxies, the Milky Way contains a population of blue, metal-poor globular clusters and another of red, metal-rich ones. Most of the latter belong to the bulge, and therefore they are poorly studied compared to the blue (halo) ones because they suffer higher extinction and larger contamination from field stars. These intrinsic difficulties, together with a lack of low-mass bulge globular clusters, are reasons to believe that their census is not complete yet. Indeed, a few new clusters have been confirmed in the last few years. One of them is VVV CL001, the subject of the present study. 
We present a new spectroscopic analysis of the recently confirmed globular cluster VVV CL001, made by means of MUSE@VLT integral field data. 
Individual spectra were extracted for stars in the VVV CL001 field. Radial velocities were derived by cross-correlation with synthetic templates. Coupled with PMs from the VVV survey, these data allow us to select 55 potential cluster members, for which we derive metallicities using the public code {\tt The Cannon}. 
The mean radial velocity of the cluster is $\mathrm{V_{helio}=-324.9 \pm 0.8 \ km \ s^{-1}}$, as estimated from 55 cluster members. This high velocity, together with a low metallicity $\mathrm{[Fe/H]=-2.04 \pm 0.02 \ dex}$ suggests that VVV CL001 could be a very old cluster. The estimated distance is $d=8.23 \pm 0.46$ kpc, placing the cluster in the Galactic bulge. Furthermore, both its current position and the orbital parameters suggest that VVV CL001 is most probably a bulge globular cluster.
\end{abstract}

\begin{keywords}
Galaxy: bulge -- Globular \ Clusters: individual: VVV CL001 -- Surveys -- Proper Motions -- Stars: abundances
\end{keywords}


\section{Introduction}

Globular clusters (GCs) offer fundamental clues to understand galaxy formation and evolution since they are the most ancient stellar systems in our Galaxy. We know that the Milky Way (MW) hosts at least 170 GCs separated between halo, bulge and thick disk GCs \citep{vasiliev2021,harris10}; however, there are arguments to think that we have not found all the bulge GCs yet. For example, comparing with M31, the MW has an important lack of GCs \citep{Caldwell2016}. The observed asymmetry in their spatial distribution around the Galactic plane suggests that some additional GCs might still be undetected, in particular those hidden inside or beyond the bulge as a result of the stellar crowding and the significant and patchy interstellar extinction \citep{baumgardt19}. Several GC candidates have been found in the last few years from the exploitation of large photometric and astrometric surveys \citep[][]{minniti11a,monibidin11,borissova14,minniti17a,minniti17b,minniti17c,Gran2019,palma19,garro20, minni21a-sgr, minni21b-sgr, garro21-sgr}. However, to confirm a GC candidate it is necessary to demonstrate that the stars have
a coherent motion, which is only possible by obtaining either the radial velocities (RV) or the proper motions (PMs), or both, for a large number of stars. To date, only a few of the candidates have been kinematically confirmed, while the great majority have been discarded as mere high-density fluctuations \citep{Koch2017, Gran2019}. 

One of the confirmed new GCs is VVV CL001 \citep{Minniti2011}, which has been discovered in the bulge region as a faint low-mass GC by the VISTA Variables in the Vía Láctea (VVV) near-IR survey \citep{vvv}. It is located at $(\ell,b)=(5.268^\circ,0.778^\circ)$ only 8 arcmin away from the known GC UKS-1. In that study, they proposed that due to the closeness of the clusters it could be the first binary GC in our Galaxy.

In the work by \cite{Gran2019}, VVV CL001 was confirmed as a GC using VVV PMs \citep[][Contreras Ramos et al. in prep.]{Contreras2017}. More recently, the study of \cite{fernandeztrincado2021} (hereafter F21) independently confirmed VVV CL001 as a GC by radial velocities, using spectra for 2 cluster stars from the Apache Point Observatory Galactic Evolution Experiment \citep[APOGEE-2S,][]{APOGEE2s}. They also found that VVV CL001 is a very metal-poor GC with a metallicity of $\rm [Fe/H]=-2.45 \pm 0.24$ dex. This metallicity contrasts with that of other bulge GCs, suggesting that the cluster might be very old. Nevertheless, they only analysed 2 stars, thus more spectra are necessary to find additional potential cluster members, and better constrain the cluster parameters. Indeed, if VVV CL001 belongs to the bulge with this metallicity, it could be the oldest GC found in this Galactic component so far. 

The paper is organized as follows: Section 2 describes the observations, the data processing and the extraction of the spectra. In section 3 we derive the radial velocities of stars in the cluster field. In section 4 we obtained the PMs and the near-IR photometry for cluster stars. Section 5 presents the metallicity determination for VVV CL001. In section 6 we estimated the distance to the cluster and obtained the orbital parameters. Finally, in section 7 we present our summary.

\section{Observations and data processing}

\subsection{Observations and calibrations}

The GC VVV CL001 was observed with the integral field spectrograph MUSE@ESO/VLT \citep{Bacon2010} on August 2017, as part of the science verification campaign \citep{museSV17} for the Ground Layer Adaptive Optics (GLAO) mode provided by the GALACSI system \citep{GALACSI}. 
The observations were performed by using the WFM\--AO mode and the nominal filter, which together provides $1\times 1$ arcmin$^2$ field-of-view (FoV), a pixel scale of $\mathrm{0.2 \ arcsec}$, and a mean spectral resolution of $\mathrm{R \sim 3000}$ over the 480\,nm\-–930\,nm range\footnote{When the WFM\--AO mode is selected the spectral coverage is not continuous because of the presence of a Na notch filter blocking the light around $\sim 590$ nm, where the spectra are dominated by the Na laser emissions. Therefore, the final spectra coverage yielded by the WFM\--AO\--Nominal setup is 480\,nm\--582\,nm and 597\,nm\-- 930\,nm.}.
The total integration time on target of 2250\,s has been achieved by combining 5 sub-exposures, each 750\,s long, with a small dithering pattern and 90$^\circ$ rotations, in order to optimize the cosmic rejection and obtaining a combined cube with uniform noise properties.

\begin{figure}
\begin{center}
	\includegraphics[width=\columnwidth]{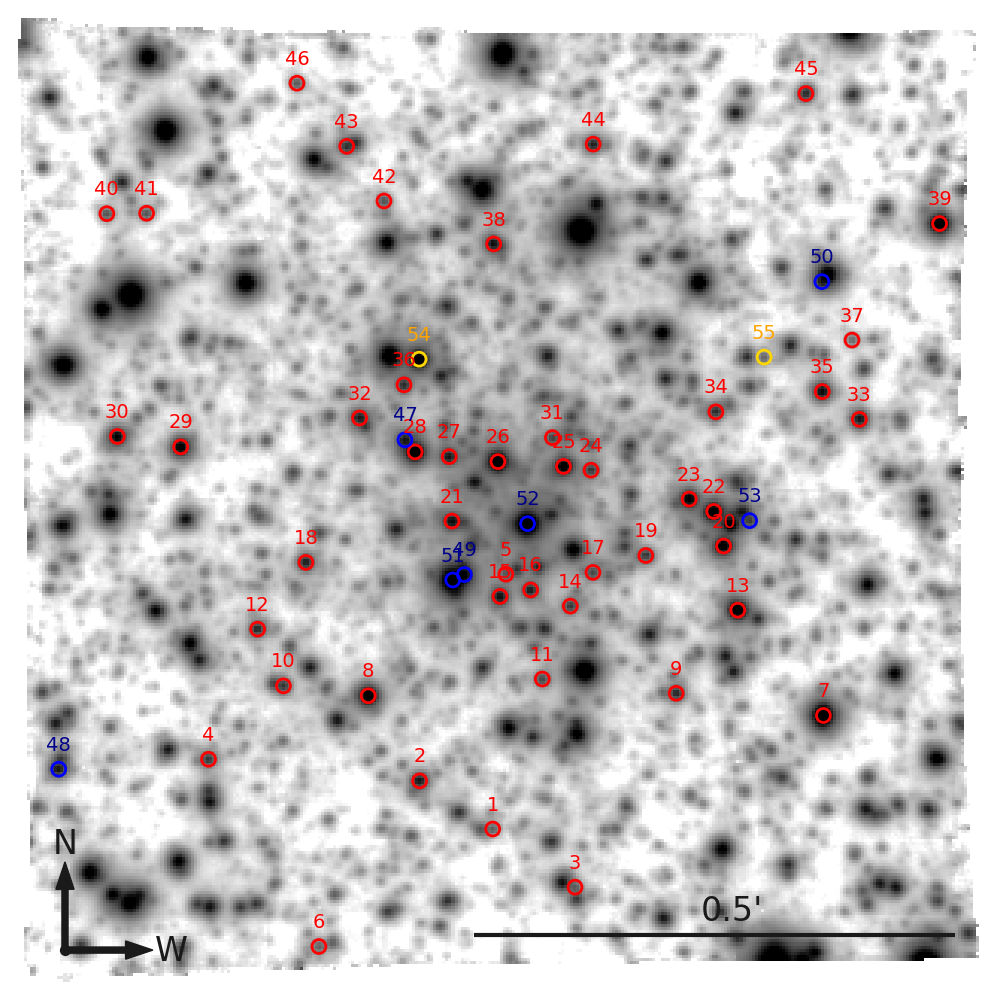}
    \caption{Members of the VVV CL001 cluster in the MUSE FoV I-band image using logarithmic scale. The red circles represent the stars that have radial velocity, PMs and metallicity measurements; the blue circles correspond to stars that only have radial velocity and metallicity; the yellow circles are stars with radial velocity and PM  measurements, but no metallicity.}
    \label{fig:cluster}
\end{center}
\end{figure}

The raw data were processed with the MUSE pipeline \citep{Weilbacher2012}, which allows to remove the instrumental signatures (e.g., bias, bad pixel map, flat-fielding, illumination corrections), to combine the individual sub\--exposures and to find the wavelength solution.
In addition, the pipeline outputs the so\--called FoV images, obtained by convolving the final MUSE datacube with the transmission curve of any standard or user\--provided filters. 
In this work we produced FoV images in the R and I\--Cousins passbands, the latter shown in Fig. \ref{fig:cluster}.

\subsection{Extraction of the spectra}

The R and I FoV images were used to detect all the point sources and perform Point Spread Function (PSF) photometry on them. The resulting master list of stars, was used to perform PSF photometry on each 
monochromatic image, obtained by slicing the final MUSE datacube along the wavelength axis.
Many of the stars in the master list are already very faint in the FoV images, and therefore their spectrum is useless as the S/N is too low. 
Nonetheless, given that the cluster center is relatively dense, detection of all the faint stars is important to de-blend the profile of the brighter ones, and therefore extract a cleaner spectrum for them.

Source detection and PSF photometry was performed by means of the DAOPHOT-II/ALLSTAR \citep{stetson+87} suite of codes. The FoV image in the I-band was analysed first, and then the output list of stars was used as input for the photometry on the FoV image in the R-band. 
About 60 bright and isolated stars were selected in the I band FoV image, and used to construct the PSF model in both the I and R FoV images. A colour magnitude diagram (CMD) was constructed by cross matching stars in both FoV images, and keeping only the 824 having photometric errors $<0.13$ mag, $\chi$ parameter $<2$ and $-1<{\rm Sharpness}<1$.

\begin{figure}
	\includegraphics[width=\columnwidth]{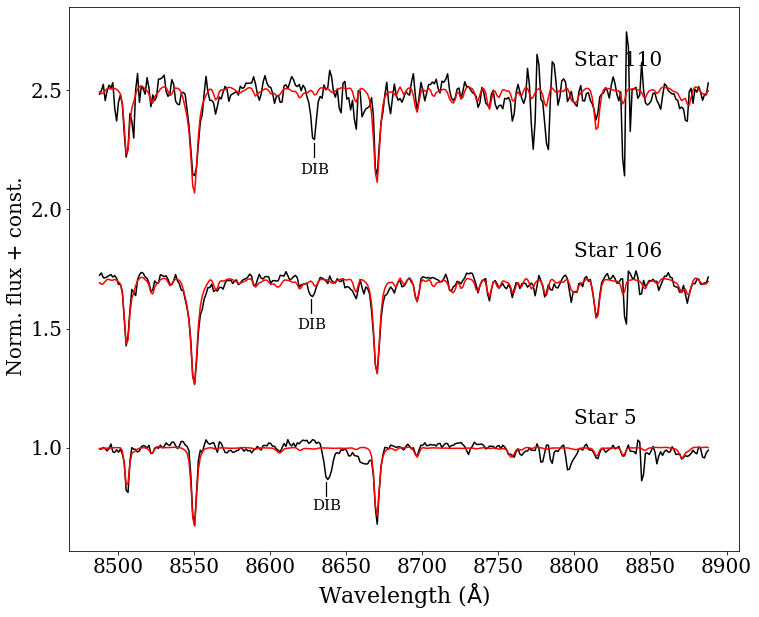}
    \caption{Randomly selected spectra for three stars within the MUSE field. Black is the observed, rest-frame spectrum, while red is the synthetic spectrum used for cross-correlation. Observed spectra were normalized and trimmed for the CaT spectral region.}
    \label{fig:spectra}
\end{figure}

These stars were then measured in each of the 800 monochromatic images corresponding to wavelengths in the range 8000-9000 \AA, i.e., the Calcium II Triplet (CaT) region. The same set of isolated and bright stars selected on the I image was blindly used to create a PSF model in each of the monochromatic images. Very importantly, the PSF fitting code ALLSTAR was set to not redetermine the stellar centroids while fitting the monochromatic images. This turned out to be crucial, because occasionally stars would almost disappear in the monochromatic images corresponding to the core of strong absorption lines. If allowed to move the centroid, ALLSTAR would often move to a nearby stars, in order to try and fit the model PSF. 
In addition, by keeping the stellar centroids fixed, we made sure that the deblending of stellar groups was done based on their position in the FoV images, where the S/N is higher than in any monochromatic one. 
A spectrum was finally reconstructed, for each star in the master list, by converting to fluxes the magnitude of the star in each monochromatic image. When magnitudes are obtained via PSF fitting with this suite of codes, it is not unusual that a star is lost, because the fit of the PSF model did not converge. In this case its magnitude is set to 99,999 mag, which translates into a {\it hole}/bad-pixel in the spectrum. This happens, however, exclusively for the faintest stars, and it is largely compensated by the fact that the spectrum for all the other stars (and even for the faint stars at other wavelengths) has a higher S/N. For this reason, after several tests, PSF fitting photometry was preferred over aperture photometry. 

In order to obtain radial velocities, as explained below, we worked around the CaT region. For this reason, each spectrum was first continuum normalized, and then trimmed to a wavelength range between 8400 \AA \ and 8800 \AA . These two steps were performed with the
\textit{continuum} and \textit{scopy} routines within the IRAF package\footnote{IRAF (Image Reduction and Analysis Facility) is distributed by the National Optical Astronomy Observatories, which are operated by the Association of Universities for Research in Astronomy, Inc., under contract with the National Science Foundation.}. Only 277 stars have spectra with well defined CaT lines, sufficient for RV determination.
Figure \ref{fig:spectra} shows three sample spectra (black lines) together with the template spectra (red lines) used in the radial velocity determination. The CaT absorption lines are clearly visible, together with an absorption band, on the left side of the reddest Ca line, that corresponds to a diffuse interstellar band (DIB). This is a well known DIB at 8620 \AA, nowadays known as the Gaia DIB \citep[][]{Geary1975,Kos2013}.

\section{Radial Velocities}

\begin{figure}
	\includegraphics[width=\columnwidth]{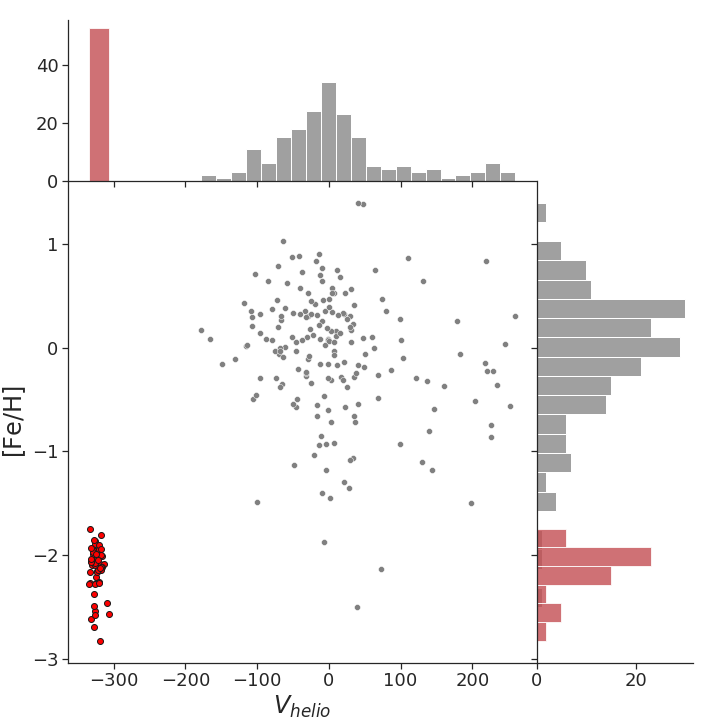}
    \caption{Metallicity versus heliocentric radial velocity of all the stars in the MUSE field. Cluster members (red dots) are very well separated from field stars (grey dots). The histograms at the top and on the right side show the projections of the distribution on the radial velocity and metallicity axis, respectively.}
    \label{fig:FEHvsRV}
\end{figure}

Radial velocities were obtained by cross-correlation using a python code written by one of us (ARA). A grid of synthetic templates was created with the SPECTRUM code \citep{Gray1994} in the CaT region, adopting model atmospheres from the APOGEE-ATLAS9 database \citep{Meszaros2012}. The grid covers a range of effective temperatures $\mathrm{T_{eff}}=$ 4000, 4250, 4500 and 4750 K; surface gravities $\mathrm{\log(g)}=$ 1.5, 2.0 and 2.5; and metallicities $\mathrm{[Fe/H]}=$ $-$2.0, $-$1.5, $-$1.0 and $-$0.5 dex, typical of giant stars in GCs. The code selects the best template for each observed spectrum in the CaT region and obtains a radial velocity by cross-correlation. Additionally, the code outputs the rest-frame corrected observed spectrum.

The distribution of heliocentric radial velocities is shown in the top histogram of Fig.~\ref{fig:FEHvsRV}. We found an isolated peak at around V$\rm _{ helio}\sim-325 \ km \ s^{-1}$, where we do not expect any contribution from bulge field stars. The 55 stars around this velocity are therefore classified as potential cluster members, currently approaching the Sun at a very large speed. The mean radial velocity of the cluster is $\mathrm{V_{helio}=-324.9 \pm 0.8 \ km \ s^{-1}}$, so we confirm the result of F21, where they conclude that the high velocity of the cluster discards the idea of the binary system between VVV CL001 and UKS-1. The peak observed at V$\rm_{helio}\sim200 \ km \ s^{-1}$ also called our attention, as a possible second cluster in the same field. The PMs of these stars, however, together with their metallicities shown in Fig.~\ref{fig:FEHvsRV} lead us to discard this possibility. Nevertheless, this peak could be related with the one obtained by \citet{Nidever2012} in the MW bulge region. Using APOGEE spectra for $\sim 4700$ K/M-giant stars they found a high-velocity peak of $\rm V_R\sim200 \ km \ s^{-1}$ for around the $10\%$ of the sample. Those stars are thought to be associated with the Galactic bar potential.

\section{Proper Motions}

\begin{figure}
	\includegraphics[width=\columnwidth]{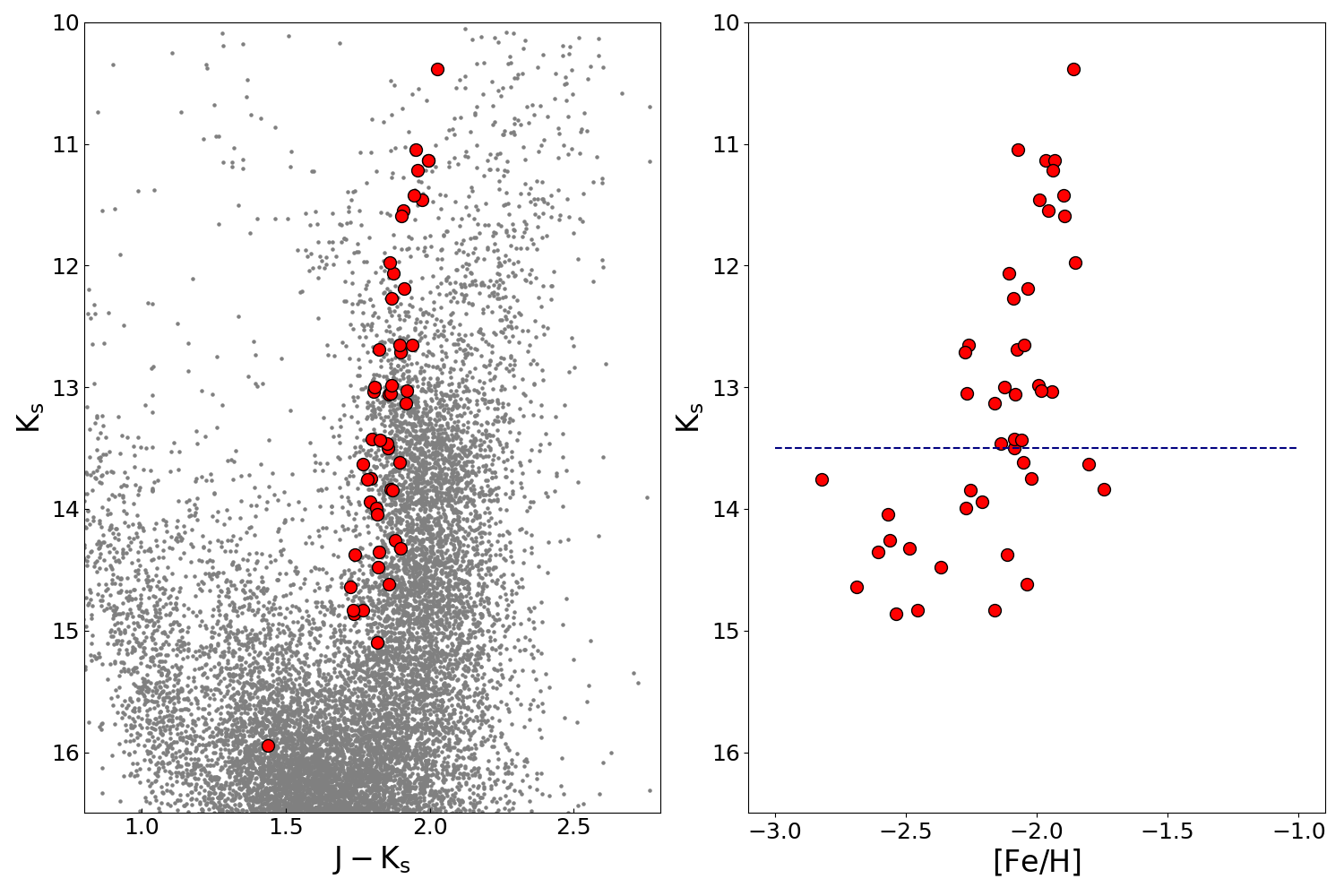}
    \caption{Left Panel: CMD for the stars detected in VVVX within 3 arcmin from the cluster center. Red symbols show the 48 cluster members, for which we have both RVs and PMs. Right Panel: Metallicity vs K$_s$ magnitude for cluster members. The dispersion in metallicity becomes significantly larger for K$_s>13.5$ mag (blue dashed line). The cluster star at K$_s\sim 16$ mag was too faint
    to allow us to derive a metallicity.}
    \label{fig:CMD}
\end{figure}

\begin{figure}
	\includegraphics[width=\columnwidth]{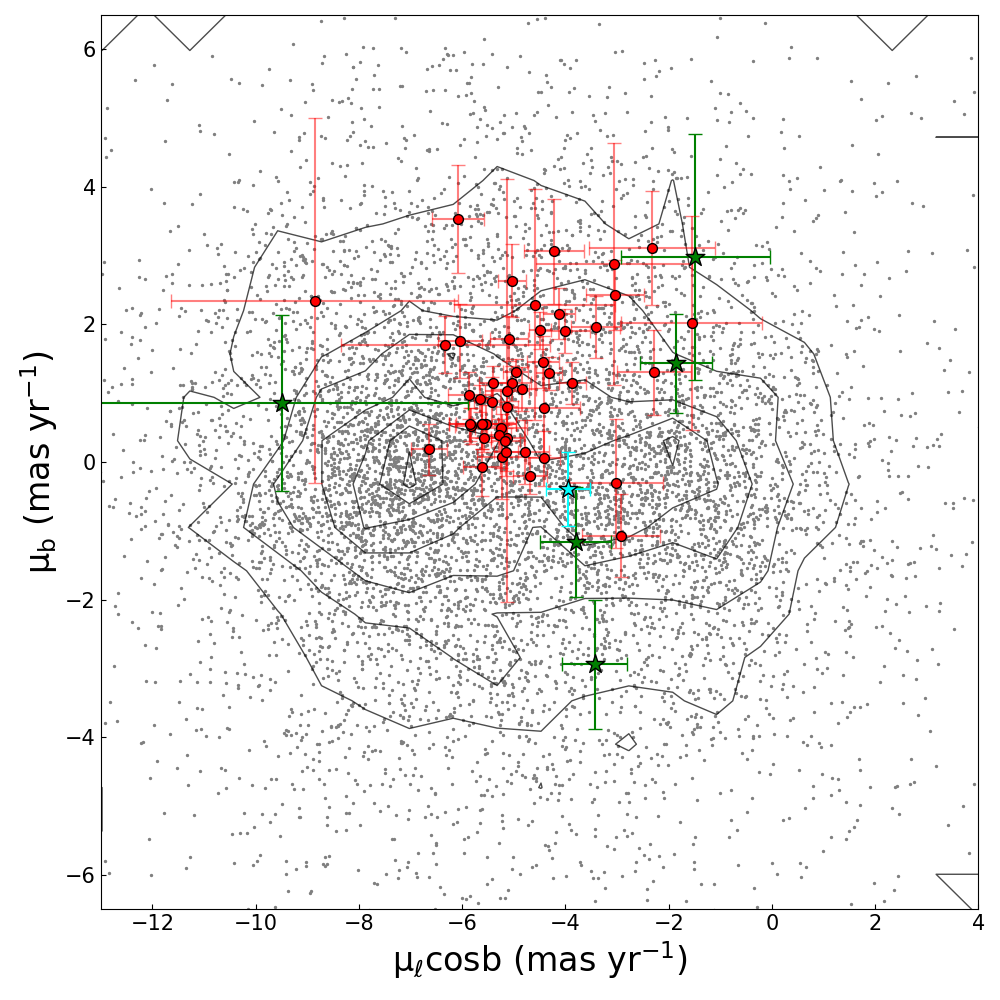}
    \caption{Vector Point Diagram for the stars detected in VVVX within 3 arcmin from the cluster center and with K$_s<16$ mag (grey dots) and for cluster members (red dots). The PMs of the RRLs (green stars) and the CepII star (cyan star) found close to the cluster center are included. Grey contours indicate the kinematic distribution with two peaks one on the left (bulge) and one on the right (disk).}
    \label{fig:VPD}
\end{figure}

Near-IR PSF photometry and PMs for stars in the MUSE field were obtained from the VVV Survey \citep{minniti10}, in the same way described by \citet{Contreras2017}. A few extra epochs were recently added to the survey, as part of its eXtension \citep[VVVX;][]{minniti2016}: these were analysed together with the original ones, yielding a 10 year baseline, from 2010 to 2019. Furthermore, the PMs were calibrated to the Gaia astrometric system to obtain the absolute PMs. Fig.~\ref{fig:CMD} shows the CMD derived for a region of 3 arcmin radius centered in VVV CL001 (in tile b351 of VVV). It includes only 48 of the 55 cluster members selected by RVs, here shown in red, as the other 7 do not have measured PMs. That is because 3 stars are saturated in VVVX, while the other 4 are too faint
and/or crowded to measure reliable PMs (see Fig.~\ref{fig:cluster}). As expected, cluster members trace a tight vertical sequence, consistent with the RGB of a metal-poor population. 

Figure~\ref{fig:VPD} shows the Vector Point Diagram (VPD) for the stars in the CMD of Fig.~\ref{fig:CMD}. Most of the 
cluster members have very small PM errors and are strongly clustered in this plot. Fainter stars, however,
have progressively higher astrometric errors and their PMs lie further away from the cluster mean value of $(\mu_l\cos{b},\mu_b)=(-4.82\pm0.12,1.14\pm0.24) \ \mathrm{mas \ yr^{-1}}$. Nonetheless, because of the extreme RVs of these stars, and since their PMs are consistent with the cluster mean motion, within their (larger)
errors, we keep them as cluster members.

\section{Metallicities using The Cannon}

The iron abundances of all the stars within the FoV of MUSE were derived using \texttt{The Cannon} \citep{Ness2015}, a data-driven approach for determining physical parameters (in this context called ``stellar labels'') from spectroscopic data. The code learns, from the known labels of a set of reference stars, how the continuum-normalized shape of the spectra depends on these labels by fitting a flexible model (with a typical quadratic polynomial model) at each wavelength. Then, The Cannon uses this model to derive labels for any set of unlabelled data, as far as it is homogeneous with respect to the reference set (i.e., same resolution and wavelength sampling). In this case, the unlabelled set of stars are our 277 stars in the MUSE field. To train our model we use a set of spectra for 3177 stars, from the Gaia ESO Survey Data Release 3 (GES DR3)\footnote{https://www.gaia-eso.eu/data-products/public-data-releases/gaia-eso-survey-data-release-3}, adopting as labels the GES recommended set of fundamental parameters (obtained by HR10+HR21 setups from GIRAFEE). The spectra were obtained with the HR21 setup ($8484-8900\ $\AA) which is centered around the CaT spectral region. The resolution of HR21 is $\mathrm{R\sim18000}$, therefore, it was degraded to the resolution of MUSE ($R\sim3000$) by a convolution with an appropriate Gaussian kernel. Then, the sampling was changed from $\Delta \lambda = 0.125$\ \AA \ to a new one $\Delta \lambda = 1.25$\ \AA \ using a python code that interpolates the wavelength points with a cubic spline in order to obtain the sampling of MUSE spectra. Also, the MUSE and the GES spectra were trimmed to a common wavelength range.

We performed a standard validation test in order to test the reliability of The Cannon to estimate stellar parameters at the resolution of MUSE given our training set. A random $10\%$ of the stars from our GES DR3 reference sample was analysed, as if it was our "science set", using the remaining $90\%$ as "training set". The procedure was repeated 10 times. A good correlation was found between the {\it known} values of the labels ($\mathrm{T_{eff},\log{g}, \ \& \ [Fe/H]} $) and those assigned by \texttt{The Cannon}, for the sample defined as "science". Specifically, the difference between the known and the assigned metallicities is, on average, consistent with zero, with a standard deviation of 0.15 dex. The spread for the effective temperature is of 165 K, around a mean bias of 27 K, while the spread for surface gravity is 0.29 dex, centered on 0.05 dex. Plots for these differences are shown in the Appendix A.

The above test demonstrates that the precision of the metallicity measurements obtained with The Cannon in this context is quite good (0.15 dex) with a negligible systematic, and therefore we measured effective temperature, surface gravity and metallicity for all stars in the MUSE field. The result is shown in Fig.~\ref{fig:FEHvsRV}. In addition to the expected broad distribution for field stars, roughly centered at [Fe/H]=0, there is a peak around [Fe/H]$\sim -2$ dex. Not surprisingly, the scatter plot in the middle panel demonstrates that this peak is due to cluster stars, with $\mathrm{V_{helio}=-324.9 \pm 0.8 \ km \ s^{-1}}$. The spread in [Fe/H] for cluster stars, in this plot, is significantly larger than what we would expect. However, if we look at the behavior of the metallicity as a function of K$_s$ magnitude (Fig.~\ref{fig:CMD}, right panel), we note that the spread in [Fe/H] increases significantly for stars fainter than K$_s$=13.5 mag. We, therefore, derive a mean cluster metallicity of [Fe/H]=$-2.04\pm0.02$ dex, based only on stars brighter than this cutoff (blue line in Fig.~\ref{fig:CMD}) where the dispersion of our measurements is consistent with the typical uncertainty in metallicity we found above.

Our mean metallicity result for VVV CL001 contrasts with the more metal-poor one of [Fe/H]=$-2.45\pm0.24$ dex from F21. Our value is based on spectra of 28 cluster members, therefore with larger statistical significance, while the value in F21 is based on two stars, although observed with high resolution spectroscopy ($R \sim 20000$). The two stars from F21 are also present in our sample (stars CL001-51 and 52 in Table \ref{table:tab1}), and they are both more metal-rich, in our measurements, by about 0.4 dex. Given the systematic error of 0.24 quoted by F21, the discrepancy between our results and theirs is lower than 2 sigmas, therefore compatible within the errors. In any case, VVV CL001 is a cluster with metallicity below -2 dex positioning it as one of the most (if not {\it the} most) metal-poor GCs in the inner Galaxy.  

\section{CL001 Orbit}

Only a small fraction of the total confirmed GCs have had their orbital parameters analyzed \citep{PerezVillegas2020}. In particular, several bulge GCs are lacking the spectroscopic follow-up observations needed to obtain their radial velocities, and therefore their orbits. An accurate distance is also a fundamental ingredient. In order to constrain it for VVV CL001, we searched for variable stars that could possibly belong to the cluster. Our search is described below, although we anticipate that we could not find any clear member of the cluster among the variables detected close to its center.

\subsection{Searching for variable stars}

The search for variable stars close to the cluster center was restricted to RR Lyrae (RRL) and type II Cepheids (CepII). We searched in the VVV catalog of variable stars available to our group (Contreras-Ramos, private communication) and found 6 RRL and 1 CepII candidates within 6 arcmin from the cluster center (Table~\ref{table:tab3}).

In order to assess their membership, we first retrieved their PMs from our VVV PMs catalog: they all had measured PM except for 1 RRL (none of these stars has MUSE spectra, since they are vary faint). Fig \ref{fig:VPD} show the PMs of these stars, together with the spectroscopically confirmed cluster members and field stars. Unfortunately, none of the variables can be considered member of the cluster with large confidence, although a couple of them have such large PM errors that they are compatible with the cluster mean PM. 

The metallicity of the RRL variables was calculated as a second attempt to establish membership. To that end, we used the routine 
\texttt{PyFiNeR} \citep{Hajdu2018}, which implements the RRL near-IR (in this case, K$_s$) light curve fitting techniques to derive robust mean magnitudes. Then, we employed the \texttt{PyMERLIN} routine \citep{Hajdu2018} in order to estimate the iron abundance of the RRLs from their K$_s$ light curve parameters. The derived metallicities, in the scale of \cite{Carretta2009}, are listed in Table \ref{table:tab3}.

\noindent
In order to estimate their distances, we use the distance modulus equation:
\begin{equation}
K_s - M_{K_s} -A_{K_s}=5 \log d-5,
\end{equation}

\noindent
where $K_s$ is the mean magnitude, $M_{K_s}$ is the absolute magnitude, and $A_{K_s}$ is the extinction coefficient. The absolute magnitude was calculated using the Period-Luminosity-Metallicity relation (PLZ) by \cite{Muraveva2015}:
\begin{equation}
M_{K_{\rm{S}}}=-2.53 \log{P_{ab}}-0.95+0.07\mathrm{[Fe/H]},
\end{equation}

\noindent
where $P_{ab}$ is the period, and [Fe/H] the photometric metallicity derived above. For the extinction coefficient,
$A_{K_s} = E_{J-K_s} \times R_{K_s}$ we adopted $R_{K_s}=0.465$ from \citet{minniti2020}, while the reddening was computed as $E_{J-K_s} = (J-K_s)-(J-K_s)_0$, with the mean intrinsic color for RRL stars $(J-K_s)_0=0.26\pm 0.03$ from \citet{Contreras2018}. The estimated distance for each RRL variable is listed in Table \ref{table:tab3}.

Five of the six RRLs are discarded as cluster members because of their metallicities, ranging from $-$1.3 dex to $-$0.7 dex: too high compared to the cluster mean metallicity [Fe/H]=$-2.04\pm0.02$ dex. The sixth one, 408\_103161, which has a metallicity of  [Fe/H]=$-1.79 \pm 0.36$ dex, closer to the cluster one, is located at a distance of 13.1 kpc. While it is possible that the cluster is further away than the Galactic bulge, as this star would indicate, we note that the star is also the only one for which no PM could be measured. In addition, it is also 2.3 arcmin away from the cluster center, outside the MUSE FoV, therefore we cannot confirm its membership by means of kinematics. Further data would be needed in order to establish whether this variable belongs to the cluster or not. 

The distance for the CepII candidate present in the region, calculated using the period-luminosity (P-L) relation for CepII stars by \cite{Bhardwaj2017A} for $J$ and $K_s$ bands, is $d=68.08$ kpc. We can rule out that it is a member of the cluster due to its very large distance. All these values can be found in the Appendix table \ref{table:tab3}.

\subsection{Isochrone fitting}

\begin{figure}
	\includegraphics[width=\columnwidth]{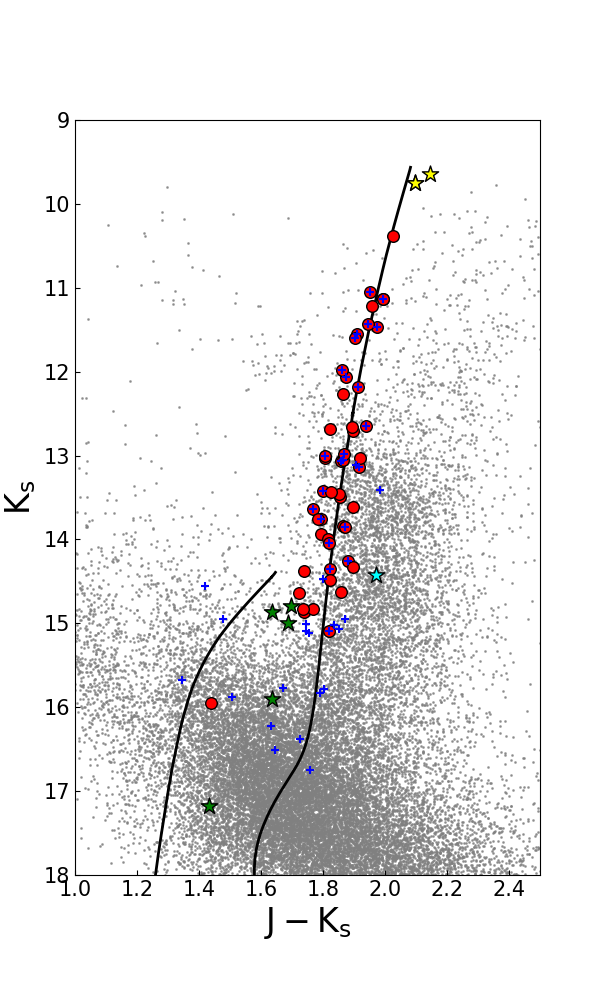}
    \caption{Isochrone fitting (black) to the observed CMD for cluster stars. Members identified here are shown in red, those derived by \protect\cite{Gran2019} are shown as blue plus signs, while the two member stars saturated in VVVX but present in 2MASS are shown as yellow stars. The RRLs (green) and the CepII candidate (cyan) are also included. Field stars from VVVX (grey dots) within 3 arcmin of the cluster center are used as background.}
    \label{fig:iso}
\end{figure}

Given that we could not derive a precise distance for the cluster by means of variable stars, we hereby obtain an approximate one 
using standard isochrone fitting to the cluster CMD. The isochrones were retrieved from the  PGPUC\footnote{http://www2.astro.puc.cl/pgpuc/index.php} web database \citep{Valcarce2012}. We need five parameters to create a PGPUC isochrone. The age is not very relevant, as our photometry does not reach the cluster main sequence turnoff, as long as it is old, as expected for a cluster of such low metallicity: we adopted 12 Gyr. The helium abundance was fixed at $Y=0.25$ considering that VVV CL001 is a very metal-poor cluster. The global metallicity was obtained by the equations available at PGPUC web page, obtaining $Z=0.00026$ using the mean cluster metallicity. For the
$\alpha$-element enhancement, we used $\rm [\alpha/Fe]=0.3$ dex as derived with APOGEE spectra, for metal-poor GCs \citep[][among others]{Schiavon2017, Meszaros2020, gran2021a}. Lastly, we adopted a standard mass loss rate of $\eta=0.2$. In addition, for the HB phase of the isochrone we have to select the HB progenitor mass, which we fixed as $\rm M_{HB}=0.8 \ M_\odot$ according the results in Figure 11 from \cite{Valcarce2012}.

Using the VVV extinction map of \cite{Surot2019} in the vicinity of VVV CL001, we found that the cluster region has a very patchy reddening $E_{J-K_s}$, varying between 1 and 1.4 mag with two peaks at 1.2 and 1.3 mag. Therefore, we kept the reddening as a free parameter, in the eyeball fitting. The best match between
the isochrone and the observed CMD is not very well constrained, given that the cluster CMD is rather vertical and featureless (Fig.~\ref{fig:iso}). Nonetheless, we could retrieve from the 2MASS catalogue \citep{2mass} the two brightest cluster members in the field, that were saturated in VVV, which gave us an idea of the location of the RGB tip. These stars were corrected to the VISTA photometric system using the equations by \cite{GonzalezFernandez2018}. In addition, we added more potential cluster candidates from \citet{Gran2019} (blue plus signs). With these constraints we could derive a distance modulus of  K$_s - M_{\rm K_s}$=15.2$\pm$ 0.1 mag, and therefore a distance of $d=8.23 \pm 0.46$ kpc, with a reddening of $E_{J-K_s}$=1.34 mag. It is fair to remember that this is an upper limit for the distance, given that the observed RGB tip could be brighter than the two brightest stars. This result agrees, within the errors, with the one derived by F21, and it places the cluster, currently, in the bulge, at the Galactic center distance.

\subsection{Orbital parameters}  

\begin{figure}
	\includegraphics[width=\columnwidth]{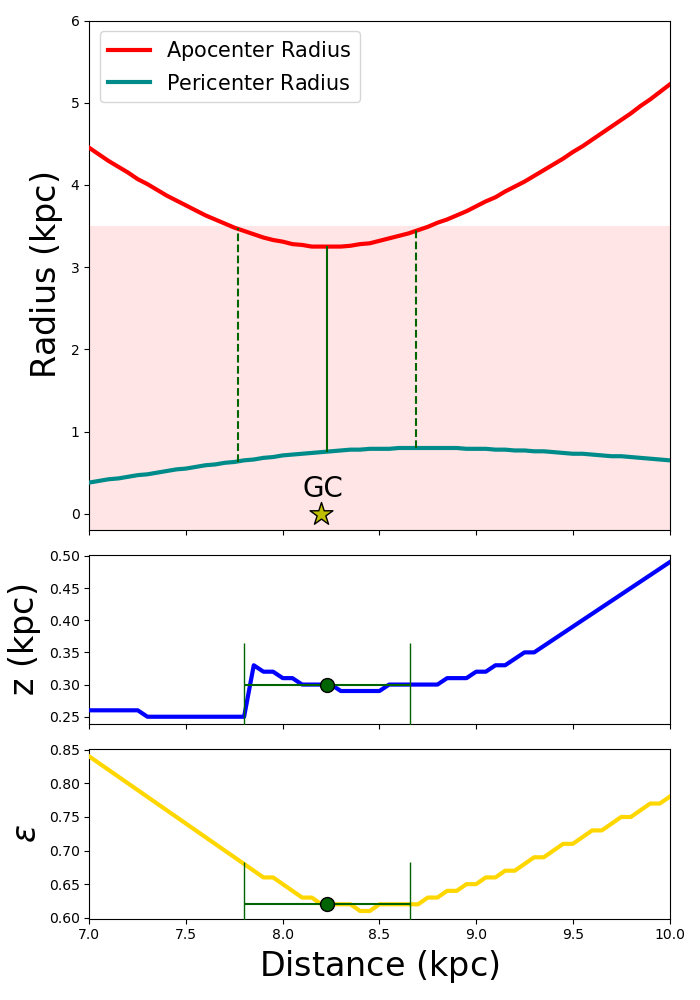}
    \caption{\textit{Top Panel:} The variation of the apocenter (red) and pericenter (dark cyan) radius for a generous range of distances to the VVV CL001 cluster. The green lines show the estimated cluster distance (solid) with its 1-sigma error (dashed).
    The region within 3.5 kpc from the Galactic center, traditionally assumed to include the bulge, is shown in pink. \textit{Middle Panel:} The height of the orbit above the plane, $z$ (blue line). The green dot is the estimated distance to VVV CL001 with its error. \textit{Bottom Panel:} The eccentricity $\epsilon$ of the cluster orbit for a range of distances.}
    \label{fig:orbits}
\end{figure}

With the approximate distance constrained above, and the three velocity components, we computed the cluster space motion and integrate its orbit. This is a qualitative analysis, whose only purpose is to investigate whether the cluster has a high probability to stay confined within the bulge, or it is just passing by. More accurate calculations will require, of course, a precise distance, which might be determined if deeper photometry is acquired and a blue horizontal branch is detected, and perhaps deeper PMs, to decontaminate it from field stars. 

The cluster radial velocity is V$_{\rm helio}=-324.9\pm0.8$ km/s and the mean absolute PM is $(\mu_l\cos{b},\mu_b)=(-4.82\pm0.12,1.14\pm0.24)$ mas/yr. The distance to the cluster was allowed to vary between 7 and 10 kpc.

In order to integrate the range of orbits of the cluster we used the package \texttt{galpy}, developed by \cite{Bovy2015}. We assumed a Solar motion with respect to the Local Standard of Rest of (U$_\odot$, V$_\odot$, W$_\odot$)=(11.1,12.24, 7.25) km/s from \cite{Schonrich2010}; a rotational velocity of the LSR of 220 km/s; and a distance of the Sun to the Galactic center of 8.2 kpc \citep{Bland2016}. The calculation requires assuming a gravitational potential for the Galaxy, in this case we used the generic potential \textit{MWPotential2014}, which includes 3 components: a spherical component (bulge) with a power-law density distribution and an exponential cut-off, a Miyamoto-Nagai disk \citep{Miyamoto1975}, and a Navarro, Frenk, and White halo potential \citep{Navarro1997}. The orbits were sampled 10000 times during a 3 Gyr time. 

In order to qualitatively analyse the range of possible orbits for VVV CL001, we examined the variation, as a function of distance, of some of the following orbital parameters: apocenter radius, pericenter radius, eccentricity, and maximum height from the Galactic plane, $z$. Figure \ref{fig:orbits} shows, in the top panel, the behavior of the apocenter (red) and pericenter (dark cyan) radius as a function of the distance. The red region represents the bulge/bar radius of $3.5$ kpc \citep{portail2017,barbuy18}. If the distance
derived here is correct within 1 sigma (i.e., between 7.8 and 8.8 kpc), then the apocenter radius is always within the bulge region, in which case we could conclude that CL001 is a bulge cluster, as it stays confined within the bulge all the time. Even if we extend the range of cluster current distance between 7 and 10 kpc, the cluster orbit would go out the bulge region only for a short time when close to the apocenter. Even in this case, however, it would spend most of the orbital time inside the bulge. 

The other orbital parameters are shown in the middle and lower panel of Fig.~\ref{fig:orbits}. The height above/below the plane, $z$, stays relatively low for the whole range of plausible distances. Nonetheless, its very low metallicity favors an association with the bulge or inner halo, rather than with the thick disk.

\section{Summary}

MUSE integral field data centred on the VVV CL001 cluster were used to measure RVs for 277 stars, of which 55 were found to be cluster members. 48 of them also have PM measured from VVV PSF photometry, and for 53 we could derive metallicities by means of The Cannon.

The heliocentric radial velocity of the cluster is V$_{\rm helio}=-324.9 \pm 0.8$ km/s, similar to previous results, based on only two member stars. This confirms VVV CL001 as one of the fastest star clusters approaching the Solar System. The high cluster velocity discards the idea of the binary system between VVV CL001 and UKS-1 proposed in previous studies. 

The metallicity was obtained by means of The Cannon, using GES iDR3 spectra smoothed to MUSE resolution as the training set. The metallicity of the cluster is [Fe/H]=$-2.04 \pm 0.02$ dex, highlighting VVV CL001 as one of the most metal-poor bulge GCs.

A distance of d=$8.23 \pm 0.46$ kpc was estimated using isochrones. Despite the relatively large error, due to the VVV photometry not reaching the sub giant branch and poorly sampling the blue horizontal branch, this distance allowed us to constrain the cluster orbit, and conclude that for all plausible distances, the cluster spends most of his time confined within the bulge region.

\section*{Acknowledgements}

J.O.C. acknowledges support from the National Agency for Research and Development (ANID) Doctorado Nacional grant 2021-21210865,
and by ESO grant SSDF21/24. Financial support for this work was also provided by the ANID BASAL Center for Astrophysics and 
Associated Technologies (CATA) through grants AFB170002, ACE210002 and FB210003, by the ANID Millennium Institute of Astrophysics (MAS) ICN12\_009 and by ANID Fondecyt Regular 1191505 (PI: M.Z.). E.V. acknowledges the Excellence Cluster ORIGINS Funded by the Deutsche Forschungsgemeinschaft (DFG, German Research Foundation) under Germany's Excellence Strategy – EXC-2094 – 390783311.
A.R.A. acknowledges support from FONDECYT through grant 3180203. F.G. acknowledges support from the CONICYT/ANID-PCHA Doctorado Nacional 2017-21171485.

Based on observations collected at the European Southern Observatory under ESO programmes 0103.D\-0386(A), 105.20MY.001, 179.B\-2002, and 198.B\-2004.

This research has made use of data obtained from the GES Data Archive, prepared and hosted by the Wide Field Astronomy Unit, Institute for Astronomy, University of Edinburgh, which is funded by the UK Science and Technology Facilities Council. 

We gratefully acknowledge the use of data from the VVV ESO
Public Survey program ID 179.B-2002 taken with the VISTA telescope, and data products from the Cambridge Astronomical Survey Unit (CASU). The VVV Survey data are made public at the ESO Archive. Based on observations taken within the ESO VISTA Public
Survey VVV, Program ID 179.B-2002.

This publication makes use of data products from the Two Micron All Sky Survey (2MASS), which is a joint project of the University of Massachusetts and the Infrared Processing and Analysis Center/California Institute of Technology, funded by the National Aeronautics and Space Administration and the National Science Foundation. 

It also made use of the NASA’s Astrophysics Data System and of the VizieR catalogue access tool, CDS, Strasbourg, France \citep{simbad}.  The original description of the VizieR service was published in \citep{vizier}. 
Finally, we acknowledge use of the following publicly available softwares: The Cannon \citep{Ness2015}, PyMERLIN: Metallicity Estimate of Rr Lyrae In the Near-infrared \citep{Hajdu2018}, PyFiNeR: Fitting Near-infrared RR Lyrae light curves \citep{Hajdu2018}, TOPCAT \citep{topcat}, pandas \citep{pandas}, IPython \citep{ipython}, numpy \citep{numpy}, matplotlib \citep{matplotlib}, Astropy, a community developed core Python package for Astronomy \citep{astropy1,astropy2}, galpy: A Python Library for Galactic Dynamics \citep{galpy} and Aladin sky atlas \citep{aladin1, aladin2}. 

\section*{Data Availability}

This project uses data obtained with the Multi Unit Spectroscopic Explorer (MUSE) under Science Verification proposal 60.A-9182(A), (P.I.: E. Valenti), together with data from the 179.B-2002 (VVV) and 198.B-2004 (VVVX) ESO VISTA Public Surveys. All the raw and pipeline processed data used in this study are publicly available. Catalogues may be made available upon request to the authors.



\bibliographystyle{mnras}
\bibliography{cl001} 

\begin{thebibliography}{}
\makeatletter
\relax
\def\mn@urlcharsother{\let\do\@makeother \do\$\do\&\do\#\do\^\do\_\do\%\do\~}
\def\mn@doi{\begingroup\mn@urlcharsother \@ifnextchar [ {\mn@doi@}
  {\mn@doi@[]}}
\def\mn@doi@[#1]#2{\def\@tempa{#1}\ifx\@tempa\@empty \href
  {http://dx.doi.org/#2} {doi:#2}\else \href {http://dx.doi.org/#2} {#1}\fi
  \endgroup}
\def\mn@eprint#1#2{\mn@eprint@#1:#2::\@nil}
\def\mn@eprint@arXiv#1{\href {http://arxiv.org/abs/#1} {{\tt arXiv:#1}}}
\def\mn@eprint@dblp#1{\href {http://dblp.uni-trier.de/rec/bibtex/#1.xml}
  {dblp:#1}}
\def\mn@eprint@#1:#2:#3:#4\@nil{\def\@tempa {#1}\def\@tempb {#2}\def\@tempc
  {#3}\ifx \@tempc \@empty \let \@tempc \@tempb \let \@tempb \@tempa \fi \ifx
  \@tempb \@empty \def\@tempb {arXiv}\fi \@ifundefined
  {mn@eprint@\@tempb}{\@tempb:\@tempc}{\expandafter \expandafter \csname
  mn@eprint@\@tempb\endcsname \expandafter{\@tempc}}}

\bibitem[\protect\citeauthoryear{{Astropy Collaboration} et~al.,}{{Astropy
  Collaboration} et~al.}{2013}]{astropy1}
{Astropy Collaboration} et~al., 2013, \mn@doi [\aap]
  {10.1051/0004-6361/201322068}, \href
  {https://ui.adsabs.harvard.edu/abs/2013A&A...558A..33A} {558, A33}

\bibitem[\protect\citeauthoryear{{Astropy Collaboration} et~al.,}{{Astropy
  Collaboration} et~al.}{2018}]{astropy2}
{Astropy Collaboration} et~al., 2018, \mn@doi [\aj] {10.3847/1538-3881/aabc4f},
  \href {https://ui.adsabs.harvard.edu/abs/2018AJ....156..123A} {156, 123}

\bibitem[\protect\citeauthoryear{{Bacon} et~al.,}{{Bacon}
  et~al.}{2010}]{Bacon2010}
{Bacon} R.,  et~al., 2010, in Ground-based and Airborne Instrumentation for
  Astronomy III. p. 773508, \mn@doi{10.1117/12.856027}

\bibitem[\protect\citeauthoryear{{Barbuy}, {Chiappini}  \& {Gerhard}}{{Barbuy}
  et~al.}{2018}]{barbuy18}
{Barbuy} B.,  {Chiappini} C.,   {Gerhard} O.,  2018, \mn@doi [\araa]
  {10.1146/annurev-astro-081817-051826}, \href
  {https://ui.adsabs.harvard.edu/abs/2018ARA&A..56..223B} {56, 223}

\bibitem[\protect\citeauthoryear{{Baumgardt}, {Hilker}, {Sollima}  \&
  {Bellini}}{{Baumgardt} et~al.}{2019}]{baumgardt19}
{Baumgardt} H.,  {Hilker} M.,  {Sollima} A.,   {Bellini} A.,  2019, \mn@doi
  [\mnras] {10.1093/mnras/sty2997}, \href
  {https://ui.adsabs.harvard.edu/abs/2019MNRAS.482.5138B} {482, 5138}

\bibitem[\protect\citeauthoryear{{Bhardwaj}, {Macri}, {Rejkuba}, {Kanbur},
  {Ngeow}  \& {Singh}}{{Bhardwaj} et~al.}{2017}]{Bhardwaj2017A}
{Bhardwaj} A.,  {Macri} L.~M.,  {Rejkuba} M.,  {Kanbur} S.~M.,  {Ngeow} C.-C.,
   {Singh} H.~P.,  2017, \mn@doi [\aj] {10.3847/1538-3881/aa5e4f}, \href
  {https://ui.adsabs.harvard.edu/abs/2017AJ....153..154B} {153, 154}

\bibitem[\protect\citeauthoryear{{Bland-Hawthorn} \&
  {Gerhard}}{{Bland-Hawthorn} \& {Gerhard}}{2016}]{Bland2016}
{Bland-Hawthorn} J.,  {Gerhard} O.,  2016, \mn@doi [\araa]
  {10.1146/annurev-astro-081915-023441}, \href
  {https://ui.adsabs.harvard.edu/abs/2016ARA&A..54..529B} {54, 529}

\bibitem[\protect\citeauthoryear{{Blanton} et~al.,}{{Blanton}
  et~al.}{2017}]{APOGEE2s}
{Blanton} M.~R.,  et~al., 2017, \mn@doi [\aj] {10.3847/1538-3881/aa7567}, \href
  {https://ui.adsabs.harvard.edu/abs/2017AJ....154...28B} {154, 28}

\bibitem[\protect\citeauthoryear{{Boch} \& {Fernique}}{{Boch} \&
  {Fernique}}{2014}]{aladin2}
{Boch} T.,  {Fernique} P.,  2014, in {Manset} N.,  {Forshay} P.,  eds,
  Astronomical Society of the Pacific Conference Series Vol. 485, Astronomical
  Data Analysis Software and Systems XXIII. p.~277

\bibitem[\protect\citeauthoryear{{Bonnarel} et~al.}{{Bonnarel}
  et~al.}{2000}]{aladin1}
{Bonnarel} F.,  et~al., 2000, \mn@doi [\aaps] {10.1051/aas:2000331}, \href
  {https://ui.adsabs.harvard.edu/abs/2000A&AS..143...33B} {143, 33}

\bibitem[\protect\citeauthoryear{{Borissova} et~al.,}{{Borissova}
  et~al.}{2014}]{borissova14}
{Borissova} J.,  et~al., 2014, \mn@doi [\aap] {10.1051/0004-6361/201322483},
  \href {https://ui.adsabs.harvard.edu/abs/2014A&A...569A..24B} {569, A24}

\bibitem[\protect\citeauthoryear{{Bovy}}{{Bovy}}{2015a}]{Bovy2015}
{Bovy} J.,  2015a, \mn@doi [\apjs] {10.1088/0067-0049/216/2/29}, \href
  {https://ui.adsabs.harvard.edu/abs/2015ApJS..216...29B} {216, 29}

\bibitem[\protect\citeauthoryear{{Bovy}}{{Bovy}}{2015b}]{galpy}
{Bovy} J.,  2015b, \mn@doi [\apjs] {10.1088/0067-0049/216/2/29}, \href
  {https://ui.adsabs.harvard.edu/abs/2015ApJS..216...29B} {216, 29}

\bibitem[\protect\citeauthoryear{{Caldwell} \& {Romanowsky}}{{Caldwell} \&
  {Romanowsky}}{2016}]{Caldwell2016}
{Caldwell} N.,  {Romanowsky} A.~J.,  2016, \mn@doi [\apj]
  {10.3847/0004-637X/824/1/42}, \href
  {https://ui.adsabs.harvard.edu/abs/2016ApJ...824...42C} {824, 42}

\bibitem[\protect\citeauthoryear{{Carretta}, {Bragaglia}, {Gratton}, {D'Orazi}
  \& {Lucatello}}{{Carretta} et~al.}{2009}]{Carretta2009}
{Carretta} E.,  {Bragaglia} A.,  {Gratton} R.,  {D'Orazi} V.,   {Lucatello} S.,
   2009, \mn@doi [\aap] {10.1051/0004-6361/200913003}, \href
  {https://ui.adsabs.harvard.edu/abs/2009A&A...508..695C} {508, 695}

\bibitem[\protect\citeauthoryear{{Contreras Ramos} et~al.,}{{Contreras Ramos}
  et~al.}{2017}]{Contreras2017}
{Contreras Ramos} R.,  et~al., 2017, \mn@doi [\aap]
  {10.1051/0004-6361/201731462}, \href
  {https://ui.adsabs.harvard.edu/abs/2017A%26A...608A.140C} {608, A140}

\bibitem[\protect\citeauthoryear{{Contreras Ramos} et~al.,}{{Contreras Ramos}
  et~al.}{2018}]{Contreras2018}
{Contreras Ramos} R.,  et~al., 2018, \mn@doi [\apj] {10.3847/1538-4357/aacf90},
  \href {https://ui.adsabs.harvard.edu/abs/2018ApJ...863...79C} {863, 79}

\bibitem[\protect\citeauthoryear{{Fern{\'a}ndez-Trincado}
  et~al.,}{{Fern{\'a}ndez-Trincado} et~al.}{2021}]{fernandeztrincado2021}
{Fern{\'a}ndez-Trincado} J.~G.,  et~al., 2021, \mn@doi [\apjl]
  {10.3847/2041-8213/abdf47}, \href
  {https://ui.adsabs.harvard.edu/abs/2021ApJ...908L..42F} {908, L42}

\bibitem[\protect\citeauthoryear{{Garro} et~al.,}{{Garro}
  et~al.}{2020}]{garro20}
{Garro} E.~R.,  et~al., 2020, \mn@doi [\aap] {10.1051/0004-6361/202039233},
  \href {https://ui.adsabs.harvard.edu/abs/2020A&A...642L..19G} {642, L19}

\bibitem[\protect\citeauthoryear{{Garro}, {Minniti}, {G{\'o}mez}  \&
  {Alonso-Garc{\'\i}a}}{{Garro} et~al.}{2021}]{garro21-sgr}
{Garro} E.~R.,  {Minniti} D.,  {G{\'o}mez} M.,   {Alonso-Garc{\'\i}a} J.,
  2021, \mn@doi [\aap] {10.1051/0004-6361/202141067}, \href
  {https://ui.adsabs.harvard.edu/abs/2021A&A...654A..23G} {654, A23}

\bibitem[\protect\citeauthoryear{{Geary}}{{Geary}}{1975}]{Geary1975}
{Geary} J.~C.,  1975, PhD thesis, University of Arizona, United States

\bibitem[\protect\citeauthoryear{{Gonz{\'a}lez-Fern{\'a}ndez}
  et~al.,}{{Gonz{\'a}lez-Fern{\'a}ndez} et~al.}{2018}]{GonzalezFernandez2018}
{Gonz{\'a}lez-Fern{\'a}ndez} C.,  et~al., 2018, \mn@doi [\mnras]
  {10.1093/mnras/stx3073}, \href
  {https://ui.adsabs.harvard.edu/abs/2018MNRAS.474.5459G} {474, 5459}

\bibitem[\protect\citeauthoryear{{Gran} et~al.,}{{Gran}
  et~al.}{2019}]{Gran2019}
{Gran} F.,  et~al., 2019, \mn@doi [\aap] {10.1051/0004-6361/201834986}, \href
  {https://ui.adsabs.harvard.edu/abs/2019A&A...628A..45G} {628, A45}

\bibitem[\protect\citeauthoryear{{Gran} et~al.,}{{Gran}
  et~al.}{2021}]{gran2021a}
{Gran} F.,  et~al., 2021, \mn@doi [\mnras] {10.1093/mnras/stab1051}, \href
  {https://ui.adsabs.harvard.edu/abs/2021MNRAS.504.3494G} {504, 3494}

\bibitem[\protect\citeauthoryear{Gray \& Corbally}{Gray \&
  Corbally}{1994}]{Gray1994}
Gray R.,  Corbally C.,  1994, The Astronomical Journal, 107, 742

\bibitem[\protect\citeauthoryear{{Hajdu}, {D{\'e}k{\'a}ny}, {Catelan}, {Grebel}
   \& {Jurcsik}}{{Hajdu} et~al.}{2018}]{Hajdu2018}
{Hajdu} G.,  {D{\'e}k{\'a}ny} I.,  {Catelan} M.,  {Grebel} E.~K.,   {Jurcsik}
  J.,  2018, \mn@doi [\apj] {10.3847/1538-4357/aab4fd}, \href
  {https://ui.adsabs.harvard.edu/abs/2018ApJ...857...55H} {857, 55}

\bibitem[\protect\citeauthoryear{{Harris}}{{Harris}}{2010}]{harris10}
{Harris} W.~E.,  2010, arXiv e-prints, \href
  {https://ui.adsabs.harvard.edu/abs/2010arXiv1012.3224H} {p. arXiv:1012.3224}

\bibitem[\protect\citeauthoryear{Hunter}{Hunter}{2007}]{matplotlib}
Hunter J.~D.,  2007, \mn@doi [Computing In Science \& Engineering]
  {10.1109/MCSE.2007.55}, 9, 90

\bibitem[\protect\citeauthoryear{{Koch}, {Kunder}  \& {Wojno}}{{Koch}
  et~al.}{2017}]{Koch2017}
{Koch} A.,  {Kunder} A.,   {Wojno} J.,  2017, \mn@doi [\aap]
  {10.1051/0004-6361/201731771}, \href
  {https://ui.adsabs.harvard.edu/abs/2017A&A...605A.128K} {605, A128}

\bibitem[\protect\citeauthoryear{{Kos} et~al.,}{{Kos} et~al.}{2013}]{Kos2013}
{Kos} J.,  et~al., 2013, \mn@doi [\apj] {10.1088/0004-637X/778/2/86}, \href
  {https://ui.adsabs.harvard.edu/abs/2013ApJ...778...86K} {778, 86}

\bibitem[\protect\citeauthoryear{{Leibundgut} et~al.,}{{Leibundgut}
  et~al.}{2017}]{museSV17}
{Leibundgut} B.,  et~al., 2017, \mn@doi [The Messenger]
  {10.18727/0722-6691/5050}, \href
  {https://ui.adsabs.harvard.edu/abs/2017Msngr.170...20L} {170, 20}

\bibitem[\protect\citeauthoryear{{M{\'e}sz{\'a}ros} et~al.,}{{M{\'e}sz{\'a}ros}
  et~al.}{2012}]{Meszaros2012}
{M{\'e}sz{\'a}ros} S.,  et~al., 2012, \mn@doi [\aj]
  {10.1088/0004-6256/144/4/120}, \href
  {https://ui.adsabs.harvard.edu/abs/2012AJ....144..120M} {144, 120}

\bibitem[\protect\citeauthoryear{{M{\'e}sz{\'a}ros} et~al.,}{{M{\'e}sz{\'a}ros}
  et~al.}{2020}]{Meszaros2020}
{M{\'e}sz{\'a}ros} S.,  et~al., 2020, \mn@doi [\mnras] {10.1093/mnras/stz3496},
  \href {https://ui.adsabs.harvard.edu/abs/2020MNRAS.492.1641M} {492, 1641}

\bibitem[\protect\citeauthoryear{{Minniti}}{{Minniti}}{2016}]{minniti2016}
{Minniti} D.,  2016, in Galactic Surveys: New Results on Formation, Evolution,
  Structure and Chemical Evolution of the Milky Way. p.~10

\bibitem[\protect\citeauthoryear{{Minniti} et~al.,}{{Minniti}
  et~al.}{2010a}]{vvv}
{Minniti} D.,  et~al., 2010a, \mn@doi [\na] {10.1016/j.newast.2009.12.002},
  \href {https://ui.adsabs.harvard.edu/abs/2010NewA...15..433M} {15, 433}

\bibitem[\protect\citeauthoryear{{Minniti} et~al.,}{{Minniti}
  et~al.}{2010b}]{minniti10}
{Minniti} D.,  et~al., 2010b, \mn@doi [\na] {10.1016/j.newast.2009.12.002},
  \href {https://ui.adsabs.harvard.edu/abs/2010NewA...15..433M} {15, 433}

\bibitem[\protect\citeauthoryear{{Minniti} et~al.,}{{Minniti}
  et~al.}{2011a}]{minniti11a}
{Minniti} D.,  et~al., 2011a, \mn@doi [\aap] {10.1051/0004-6361/201015795},
  \href {https://ui.adsabs.harvard.edu/abs/2011A&A...527A..81M} {527, A81}

\bibitem[\protect\citeauthoryear{{Minniti} et~al.,}{{Minniti}
  et~al.}{2011b}]{Minniti2011}
{Minniti} D.,  et~al., 2011b, \mn@doi [\aap] {10.1051/0004-6361/201015795},
  \href {https://ui.adsabs.harvard.edu/abs/2011A%26A...527A..81M} {527, A81}

\bibitem[\protect\citeauthoryear{{Minniti}, {Alonso-Garc{\'\i}a}, {Braga},
  {Contreras Ramos}, {Hempel}, {Palma}, {Pullen}  \& {Saito}}{{Minniti}
  et~al.}{2017a}]{minniti17b}
{Minniti} D.,  {Alonso-Garc{\'\i}a} J.,  {Braga} V.,  {Contreras Ramos} R.,
  {Hempel} M.,  {Palma} T.,  {Pullen} J.,   {Saito} R.~K.,  2017a, \mn@doi
  [Research Notes of the American Astronomical Society]
  {10.3847/2515-5172/aa9ab7}, \href
  {https://ui.adsabs.harvard.edu/abs/2017RNAAS...1...16M} {1, 16}

\bibitem[\protect\citeauthoryear{{Minniti}, {Alonso-Garc{\'\i}a}  \&
  {Pullen}}{{Minniti} et~al.}{2017b}]{minniti17c}
{Minniti} D.,  {Alonso-Garc{\'\i}a} J.,   {Pullen} J.,  2017b, \mn@doi
  [Research Notes of the American Astronomical Society]
  {10.3847/2515-5172/aaa3ed}, \href
  {https://ui.adsabs.harvard.edu/abs/2017RNAAS...1...54M} {1, 54}

\bibitem[\protect\citeauthoryear{{Minniti} et~al.,}{{Minniti}
  et~al.}{2017c}]{minniti17a}
{Minniti} D.,  et~al., 2017c, \mn@doi [\apjl] {10.3847/2041-8213/aa95b8}, \href
  {https://ui.adsabs.harvard.edu/abs/2017ApJ...849L..24M} {849, L24}

\bibitem[\protect\citeauthoryear{{Minniti} et~al.,}{{Minniti}
  et~al.}{2020}]{minniti2020}
{Minniti} J.~H.,  et~al., 2020, \mn@doi [\aap] {10.1051/0004-6361/202037575},
  \href {https://ui.adsabs.harvard.edu/abs/2020A&A...640A..92M} {640, A92}

\bibitem[\protect\citeauthoryear{{Minniti} et~al.,}{{Minniti}
  et~al.}{2021a}]{minni21a-sgr}
{Minniti} D.,  et~al., 2021a, \mn@doi [\aap] {10.1051/0004-6361/202140395},
  \href {https://ui.adsabs.harvard.edu/abs/2021A&A...647L...4M} {647, L4}

\bibitem[\protect\citeauthoryear{{Minniti}, {G{\'o}mez}, {Alonso-Garc{\'\i}a},
  {Saito}  \& {Garro}}{{Minniti} et~al.}{2021b}]{minni21b-sgr}
{Minniti} D.,  {G{\'o}mez} M.,  {Alonso-Garc{\'\i}a} J.,  {Saito} R.~K.,
  {Garro} E.~R.,  2021b, \mn@doi [\aap] {10.1051/0004-6361/202140714}, \href
  {https://ui.adsabs.harvard.edu/abs/2021A&A...650L..12M} {650, L12}

\bibitem[\protect\citeauthoryear{{Miyamoto} \& {Nagai}}{{Miyamoto} \&
  {Nagai}}{1975}]{Miyamoto1975}
{Miyamoto} M.,  {Nagai} R.,  1975, \pasj, \href
  {https://ui.adsabs.harvard.edu/abs/1975PASJ...27..533M} {27, 533}

\bibitem[\protect\citeauthoryear{{Moni Bidin} et~al.,}{{Moni Bidin}
  et~al.}{2011}]{monibidin11}
{Moni Bidin} C.,  et~al., 2011, \mn@doi [\aap] {10.1051/0004-6361/201117488},
  \href {https://ui.adsabs.harvard.edu/abs/2011A&A...535A..33M} {535, A33}

\bibitem[\protect\citeauthoryear{{Muraveva} et~al.,}{{Muraveva}
  et~al.}{2015}]{Muraveva2015}
{Muraveva} T.,  et~al., 2015, \mn@doi [\apj] {10.1088/0004-637X/807/2/127},
  \href {https://ui.adsabs.harvard.edu/abs/2015ApJ...807..127M} {807, 127}

\bibitem[\protect\citeauthoryear{{Navarro}, {Frenk}  \& {White}}{{Navarro}
  et~al.}{1997}]{Navarro1997}
{Navarro} J.~F.,  {Frenk} C.~S.,   {White} S. D.~M.,  1997, \mn@doi [\apj]
  {10.1086/304888}, \href
  {https://ui.adsabs.harvard.edu/abs/1997ApJ...490..493N} {490, 493}

\bibitem[\protect\citeauthoryear{{Ness}, {Hogg}, {Rix}, {Ho}  \&
  {Zasowski}}{{Ness} et~al.}{2015}]{Ness2015}
{Ness} M.,  {Hogg} D.~W.,  {Rix} H.-W.,  {Ho} A.~Y.~Q.,   {Zasowski} G.,  2015,
  \mn@doi [\apj] {10.1088/0004-637X/808/1/16}, \href
  {https://ui.adsabs.harvard.edu/abs/2015ApJ...808...16N} {808, 16}

\bibitem[\protect\citeauthoryear{{Nidever} et~al.,}{{Nidever}
  et~al.}{2012}]{Nidever2012}
{Nidever} D.~L.,  et~al., 2012, \mn@doi [\apjl] {10.1088/2041-8205/755/2/L25},
  \href {https://ui.adsabs.harvard.edu/abs/2012ApJ...755L..25N} {755, L25}

\bibitem[\protect\citeauthoryear{{Ochsenbein} et~al.}{{Ochsenbein}
  et~al.}{2000}]{vizier}
{Ochsenbein} F.,  et~al., 2000, \mn@doi [\aaps] {10.1051/aas:2000169}, \href
  {https://ui.adsabs.harvard.edu/abs/2000A&AS..143...23O} {143, 23}

\bibitem[\protect\citeauthoryear{{Palma} et~al.,}{{Palma}
  et~al.}{2019}]{palma19}
{Palma} T.,  et~al., 2019, \mn@doi [\mnras] {10.1093/mnras/stz1489}, \href
  {https://ui.adsabs.harvard.edu/abs/2019MNRAS.487.3140P} {487, 3140}

\bibitem[\protect\citeauthoryear{P\'erez \& Granger}{P\'erez \&
  Granger}{2007}]{ipython}
P\'erez F.,  Granger B.~E.,  2007, \mn@doi [Computing in Science and
  Engineering] {10.1109/MCSE.2007.53}, 9, 21

\bibitem[\protect\citeauthoryear{{P{\'e}rez-Villegas}, {Barbuy}, {Kerber},
  {Ortolani}, {Souza}  \& {Bica}}{{P{\'e}rez-Villegas}
  et~al.}{2020}]{PerezVillegas2020}
{P{\'e}rez-Villegas} A.,  {Barbuy} B.,  {Kerber} L.~O.,  {Ortolani} S.,
  {Souza} S.~O.,   {Bica} E.,  2020, \mn@doi [\mnras] {10.1093/mnras/stz3162},
  \href {https://ui.adsabs.harvard.edu/abs/2020MNRAS.491.3251P} {491, 3251}

\bibitem[\protect\citeauthoryear{{Portail}, {Gerhard}, {Wegg}  \&
  {Ness}}{{Portail} et~al.}{2017}]{portail2017}
{Portail} M.,  {Gerhard} O.,  {Wegg} C.,   {Ness} M.,  2017, \mn@doi [\mnras]
  {10.1093/mnras/stw2819}, \href
  {https://ui.adsabs.harvard.edu/abs/2017MNRAS.465.1621P} {465, 1621}

\bibitem[\protect\citeauthoryear{{Schiavon} et~al.,}{{Schiavon}
  et~al.}{2017}]{Schiavon2017}
{Schiavon} R.~P.,  et~al., 2017, \mn@doi [\mnras] {10.1093/mnras/stw3093},
  \href {https://ui.adsabs.harvard.edu/abs/2017MNRAS.466.1010S} {466, 1010}

\bibitem[\protect\citeauthoryear{{Sch{\"o}nrich}, {Binney}  \&
  {Dehnen}}{{Sch{\"o}nrich} et~al.}{2010}]{Schonrich2010}
{Sch{\"o}nrich} R.,  {Binney} J.,   {Dehnen} W.,  2010, \mn@doi [\mnras]
  {10.1111/j.1365-2966.2010.16253.x}, \href
  {https://ui.adsabs.harvard.edu/abs/2010MNRAS.403.1829S} {403, 1829}

\bibitem[\protect\citeauthoryear{{Skrutskie} et~al.,}{{Skrutskie}
  et~al.}{2006}]{2mass}
{Skrutskie} M.~F.,  et~al., 2006, \mn@doi [\aj] {10.1086/498708}, \href
  {https://ui.adsabs.harvard.edu/abs/2006AJ....131.1163S} {131, 1163}

\bibitem[\protect\citeauthoryear{{Stetson}}{{Stetson}}{1987}]{stetson+87}
{Stetson} P.~B.,  1987, \mn@doi [\pasp] {10.1086/131977}, \href
  {https://ui.adsabs.harvard.edu/abs/1987PASP...99..191S} {99, 191}

\bibitem[\protect\citeauthoryear{{Str{\"o}bele} et~al.,}{{Str{\"o}bele}
  et~al.}{2012}]{GALACSI}
{Str{\"o}bele} S.,  et~al., 2012, in {Ellerbroek} B.~L.,  {Marchetti} E.,
  {V{\'e}ran} J.-P.,  eds,  Society of Photo-Optical Instrumentation Engineers
  (SPIE) Conference Series Vol. 8447, Adaptive Optics Systems III. p. 844737,
  \mn@doi{10.1117/12.926110}

\bibitem[\protect\citeauthoryear{{Surot} et~al.,}{{Surot}
  et~al.}{2019}]{Surot2019}
{Surot} F.,  et~al., 2019, \mn@doi [\aap] {10.1051/0004-6361/201935730}, \href
  {https://ui.adsabs.harvard.edu/abs/2019A&A...629A...1S} {629, A1}

\bibitem[\protect\citeauthoryear{{Taylor}}{{Taylor}}{2005}]{topcat}
{Taylor} M.~B.,  2005, in {Shopbell} P.,  {Britton} M.,   {Ebert} R.,  eds,
  Astronomical Society of the Pacific Conference Series Vol. 347, Astronomical
  Data Analysis Software and Systems XIV. p.~29

\bibitem[\protect\citeauthoryear{{Valcarce}, {Catelan}  \&
  {Sweigart}}{{Valcarce} et~al.}{2012}]{Valcarce2012}
{Valcarce} A.~A.~R.,  {Catelan} M.,   {Sweigart} A.~V.,  2012, \mn@doi [\aap]
  {10.1051/0004-6361/201219510}, \href
  {https://ui.adsabs.harvard.edu/abs/2012A&A...547A...5V} {547, A5}

\bibitem[\protect\citeauthoryear{{Vasiliev} \& {Baumgardt}}{{Vasiliev} \&
  {Baumgardt}}{2021}]{vasiliev2021}
{Vasiliev} E.,  {Baumgardt} H.,  2021, \mn@doi [\mnras]
  {10.1093/mnras/stab1475}, \href
  {https://ui.adsabs.harvard.edu/abs/2021MNRAS.505.5978V} {505, 5978}

\bibitem[\protect\citeauthoryear{{Weilbacher}, {Streicher}, {Urrutia}, {Jarno},
  {P{\'e}contal-Rousset}, {Bacon}  \& {B{\"o}hm}}{{Weilbacher}
  et~al.}{2012}]{Weilbacher2012}
{Weilbacher} P.~M.,  {Streicher} O.,  {Urrutia} T.,  {Jarno} A.,
  {P{\'e}contal-Rousset} A.,  {Bacon} R.,   {B{\"o}hm} P.,  2012, in \procspie.
  p. 84510B, \mn@doi{10.1117/12.925114}

\bibitem[\protect\citeauthoryear{{Wenger} et~al.}{{Wenger}
  et~al.}{2000}]{simbad}
{Wenger} M.,  et~al., 2000, \mn@doi [\aaps] {10.1051/aas:2000332}, \href
  {https://ui.adsabs.harvard.edu/abs/2000A&AS..143....9W} {143, 9}

\bibitem[\protect\citeauthoryear{pandas~development team}{pandas~development
  team}{2020}]{pandas}
pandas~development team T.,  2020, pandas-dev/pandas: Pandas,
  \mn@doi{10.5281/zenodo.3509134}, \url
  {https://doi.org/10.5281/zenodo.3509134}

\bibitem[\protect\citeauthoryear{van~der Walt \& Varoquaux}{van~der Walt \&
  Varoquaux}{2011}]{numpy}
van~der Walt S.;~Colbert S.~C.,  Varoquaux G.,  2011, \mn@doi [Computing in
  Science and Engineering] {10.1109/MCSE.2011.37}, 13, 22

\makeatother
\end{thebibliography}

\appendix

\section{Test of the performance of The Cannon}
\begin{figure*}
	\includegraphics[width=15cm]{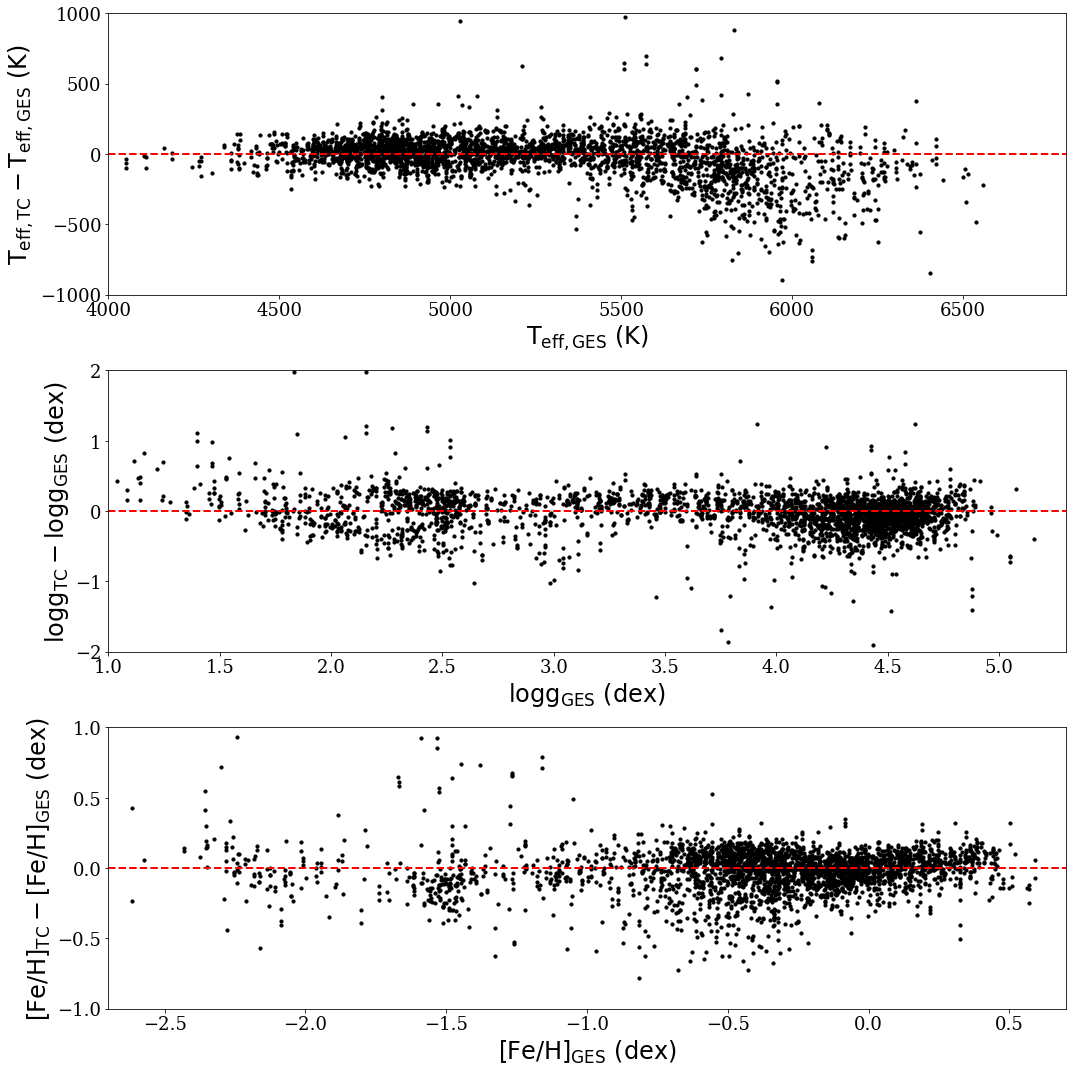}
    \caption{Difference between the stellar parameters assigned by The Cannon (TC) and those "known" from the GES iDR3 database (GES) used as the training set. Ten runs of random 10\% of the sample (described in Sec 5) were used as a test to do the verification.}
    \label{fig:params}
\end{figure*}

\section{Cluster members \& parameters}

\begin{table*}
\caption{Cluster members and their parameters. IDs of the stars, coordinates, heliocentric radial velocity, metallicity, J and  K$_s$-band photometry along with PMs from VVV were presented.}
\begin{threeparttable}
\centering
\footnotesize 
\begin{tabular}{lccccccccrr}
\hline \hline
ID &   RA   &   DEC   &  V$_{\rm helio}$  &  [Fe/H]   &  $\ell$  &  $b$  &    J    &   K$_s$  & $\mu_\ell \cos{b}$ & $\mu_b$ \\
   &  (deg) &  (deg)  &   $\rm km \ s^{-1}$          &   (dex)   & (deg) & (deg) &  (mag)  &  (mag)   & ($\rm mas \ yr^{-1}$)             & ($\rm mas \ yr^{-1}$) \\ 
\hline
VVV CL001-1  & 268.677091 & $-$24.020634 & $-$326.3 & $-$2.54 & 5.262540 & 0.776809 & 16.597 & 14.859 & $-$3.021  & $-$0.311 \\ 
VVV CL001-2  & 268.678557 & $-$24.019754 & $-$327.3 & $-$2.08 & 5.263961 & 0.776106 & 14.920 & 13.061 & $-$5.236  & 0.075  \\
VVV CL001-3  & 268.675440 & $-$24.021696 & $-$333.0 & $-$1.74 & 5.260655 & 0.777512 & 15.704 & 13.839 & $-$2.92  & $-$1.071 \\
VVV CL001-4  & 268.682793 & $-$24.019359 & $-$324.3 & $-$2.21 & 5.266260 & 0.772974 & 15.735 & 13.941 & $-$4.438 & 1.459   \\
VVV CL001-5  & 268.676827 & $-$24.015979 & $-$327.0 & $-$2.69 & 5.266422 & 0.779405 & 16.364 & 14.640 & $-$3.067 & 2.88  \\
VVV CL001-6  & 268.680578 & $-$24.022780 & $-$333.4 & $-$2.16 & 5.262282 & 0.773010 & 16.602 & 14.835 & $-$4.22  & 3.062  \\
VVV CL001-7  & 268.670456 & $-$24.018559 & $-$326.6 & $-$1.86 & 5.261257 & 0.783106 & 12.411 & 10.384 & $-$4.592 & 2.286  \\
VVV CL001-8  & 268.679588 & $-$24.018199 & $-$325.9 & $-$1.96 & 5.265780 & 0.776083 & 13.458 & 11.549 & $-$5.547 & 0.559  \\
VVV CL001-9  & 268.673410 & $-$24.018154 & $-$317.2 & $-$1.80 & 5.262969 & 0.780974 & 15.402 & 13.633 & $-$4.688 & $-$0.202 \\
VVV CL001-10 & 268.681290 & $-$24.018018 & $-$318.6 & $-$2.08 & 5.266748 & 0.774811 & 15.355 & 13.499 & $-$8.854 & 2.346  \\
\hline
VVV CL001-11 & 268.676096 & $-$24.017894 & $-$306.9 & $-$2.56 & 5.264428 & 0.778999 & 16.136 & 14.256 & $-$5.861 & 0.969  \\
VVV CL001-12 & 268.681808 & $-$24.016982 & $-$321.2 & $-$2.02 & 5.267852 & 0.774957 & 15.547 & 13.752 & $-$5.429 & 0.879  \\
VVV CL001-13 & 268.672175 & $-$24.016635 & $-$326.3 & $-$1.90 & 5.263705 & 0.782713 & 13.495 & 11.593 & $-$5.569 & 0.349  \\
VVV CL001-14 & 268.675534 & $-$24.016561 & $-$320.0 & $-$2.25 & 5.265320 & 0.780118 & 15.718 & 13.846 & $-$5.123 & 0.352  \\
VVV CL001-15 & 268.676944 & $-$24.016385 & $-$333.0 & $-$2.26 & 5.266118 & 0.779090 & 14.590 & 12.652 & $-$5.837 & 0.521  \\
VVV CL001-16 & 268.676330 & $-$24.016266 & $-$323.2 & $-$2.16 & 5.265937 & 0.779633 & 15.048 & 13.131 & $-$5.845 & 0.55  \\
VVV CL001-17 & 268.675079 & $-$24.015945 & $-$330.9 & $-$2.61 & 5.265650 & 0.780784 & 16.178 & 14.355 & $-$5.251 & 0.488  \\
VVV CL001-18 & 268.680839 & $-$24.015764 & $-$313.1 & $-$2.08 & 5.268446 & 0.776337 & 15.225 & 13.426 & $-$5.128 & 0.806  \\
VVV CL001-19 & 268.674020 & $-$24.015638 & $-$326.3 & $-$2.27 & 5.265410 & 0.781758 & 15.805 & 13.990 & $-$6.645 & 0.185  \\
VVV CL001-20 & 268.672459 & $-$24.015464 & $-$316.6 & $-$2.11 & 5.264843 & 0.783073 & 13.935 & 12.062 & $-$5.647 & 0.921  \\
\hline
VVV CL001-21 & 268.677908 & $-$24.015009 & $-$333.7 & $-$2.28 & 5.267745 & 0.779033 & 14.610 & 12.712 & $-$6.048 & 1.756  \\
VVV CL001-22 & 268.672656 & $-$24.014832 & $-$329.2 & $-$1.99 & 5.265480 & 0.783237 & 13.437 & 11.463 & $-$5.28  & 0.397  \\
VVV CL001-23 & 268.673148 & $-$24.014606 & $-$322.9 & $-$2.03 & 5.265901 & 0.782969 & 14.098 & 12.186 & $-$5.148 & 0.147  \\
VVV CL001-24 & 268.675119 & $-$24.014078 & $-$325.9 & $-$2.57 & 5.267268 & 0.781686 & 15.867 & 14.048 & $-$4.778 & 0.144  \\
VVV CL001-25 & 268.675671 & $-$24.014010 & $-$327.3 & $-$1.85 & 5.267581 & 0.781287 & 13.838 & 11.976 & $-$5.169 & 0.301  \\
VVV CL001-26 & 268.676986 & $-$24.013922 & $-$325.6 & $-$2.07 & 5.268260 & 0.780302 & 12.999 & 11.048 & $-$5.626 & $-$0.077  \\
VVV CL001-27 & 268.677963 & $-$24.013830 & $-$320.4 & $-$2.27 & 5.268795 & 0.779581 & 14.915 & 13.051 & $-$4.408 & 0.056  \\
VVV CL001-28 & 268.678648 & $-$24.013745 & $-$329.5 & $-$1.97 & 5.269188 & 0.779086 & 13.128 & 11.134 & $-$5.613 & 0.555  \\
VVV CL001-29 & 268.683352 & $-$24.013652 & $-$326.6 & $-$2.09 & 5.271435 & 0.775424 & 14.138 & 12.272 & $-$4.493 & 1.917   \\
VVV CL001-30 & 268.684622 & $-$24.013462 & $-$326.7 & $-$2.08 & 5.272180 & 0.77451  & 14.511 & 12.687 & $-$4.32  & 1.29  \\
\hline
VVV CL001-31 & 268.675893 & $-$24.013484 & $-$317.6 & $-$2.11 & 5.268137 & 0.78139  & 16.113 & 14.373 & $-$6.082 & 3.533  \\
VVV CL001-32 & 268.679762 & $-$24.013125 & $-$320.4 & $-$1.94 & 5.270195 & 0.778525 & 14.840 & 13.035 & $-$6.325 & 1.707  \\
VVV CL001-33 & 268.669735 & $-$24.013149 & $-$317.2 & $-$1.99 & 5.265584 & 0.786403 & 14.852 & 12.985 & $-$5.026 & 1.142  \\
VVV CL001-34 & 268.672614 & $-$24.013013 & $-$321.2 & $-$2.14 & 5.267028 & 0.784201 & 15.314 & 13.462 & $-$5.036 & 2.635  \\
VVV CL001-35 & 268.670479 & $-$24.012643 & $-$321.4 & $-$2.05 & 5.266369 & 0.786065 & 14.549 & 12.654 & $-$4.963 & 1.305  \\
VVV CL001-36 & 268.678871 & $-$24.012520 & $-$331.6 & $-$2.05 & 5.270342 & 0.779526 & 15.515 & 13.618 & $-$5.093 & 1.796  \\
VVV CL001-37 & 268.678567 & $-$24.012050 & $-$321.1 & $-$1.90 & 5.270604 & 0.779995 & 13.371 & 11.426 & $-$4.846 & 1.068   \\
VVV CL001-38 & 268.677071 & $-$24.009945 & $-$319.7 & $-$2.12 & 5.271728 & 0.782240 & 14.808 & 13.001 & $-$5.405 & 1.149  \\
VVV CL001-39 & 268.668125 & $-$24.009573 & $-$317.7 & $-$1.94 & 5.267928 & 0.789476 & 13.176 & 11.219 & $-$3.867 & 1.15  \\
VVV CL001-40 & 268.684832 & $-$24.009391 & $-$327.5 & $-$2.48 & 5.275787 & 0.776401 & 16.223 & 14.325 & $-$3.415 & 1.957  \\
\hline
VVV CL001-41 & 268.684035 & $-$24.009384 & $-$327.7 & $-$2.37 & 5.275434 & 0.777035 & 16.304 & 14.482 & $-$2.288 & 1.302   \\
VVV CL001-42 & 268.679273 & $-$24.009161 & $-$309.0 & $-$2.46 & 5.273416 & 0.780894 & 16.569 & 14.834 & $-$2.328 & 3.112  \\
VVV CL001-43 & 268.680020 & $-$24.008162 & $-$324.9 & $-$1.98 & 5.274616 & 0.780928 & 14.947 & 13.027 & $-$5.129 & 1.039  \\
VVV CL001-44 & 268.675076 & $-$24.008122 & $-$331.3 & $-$2.06 & 5.272381 & 0.784734 & 15.257 & 13.430 & $-$3.999 & 1.907  \\
VVV CL001-45 & 268.670808 & $-$24.007198 & $-$319.7 & $-$2.82 & 5.271201 & 0.788551 & 15.538 & 13.755 & $-$4.131 & 2.157   \\
VVV CL001-46 & 268.681018 & $-$24.007008 & $-$330.9 & $-$2.04 & 5.276085 & 0.780607 & 16.478 & 14.621 & $-$3.035 & 2.434  \\
VVV CL001-47 & 268.678852 & $-$24.013524 & $-$331.9 & $-$1.93 & 5.26933  & 0.77894  & -      & -      & -     & -       \\
VVV CL001-48 & 268.685798 & $-$24.019541 & $-$324.9 & $-$2.26 & 5.26733  & 0.77043  & -      & -      & -     & -       \\
VVV CL001-49 & 268.677660 & $-$24.015985 & $-$322.8 & $-$2.15 & 5.26664  & 0.77864  & -      & -      & -     & -       \\
VVV CL001-50 & 268.670488 & $-$24.010634 & $-$315.9 & $-$2.01 & 5.26621  & 0.78599  & -      & -      & -     & -       \\
\hline
VVV CL001-51 & 268.677893 & $-$24.016084 & $-$324.3 & $-$2.04 & 5.26666  & 0.77841  & 11.849*      & 9.751*     & -     & -       \\
VVV CL001-52 & 268.676392 & $-$24.015055 & $-$330.9 & $-$2.09 & 5.26686  & 0.78011  & 11.788*      & 9.642*      & -     & -       \\
VVV CL001-53 & 268.671937 & $-$24.015000 & $-$318.3 & $-$2.14 & 5.26485  & 0.78365  & -      & -      & -     & -       \\
VVV CL001-54 & 268.671646 & $-$24.012012 & $-$300.3 & -       & 5.267445 & 0.785473 & 17.386 & 15.947 & -1.558 & 2.023   \\
VVV CL001-55 & 268.669883 & $-$24.011701 & $-$317.3 & -       & 5.266896 & 0.787017 & 16.916 & 15.097 & -4.417 & 0.784   \\
\hline
\end{tabular}
\begin{tablenotes}
\item * Photometry obtained from 2MASS and corrected by \cite{GonzalezFernandez2018}.
\end{tablenotes}
\label{table:tab1}
\end{threeparttable}
\end{table*}

\section{Variable stars in the vecinity of VVV CL001}

\begin{table*}
\centering
\scriptsize
\caption{The parameters of the variable stars in the vicinity of VVV CL001, all of them in tile b351. VVV-ID of the stars, coordinates, variable type, period, distance to the center of the cluster, PMs and photometry from VVV, extinction coefficient and estimated distance are presented.}
\begin{tabular}{lcclcccrrcccc}
\hline \hline
VVV-ID  &  RA   & DEC   & Type & P   & [Fe/H]   & R  & $\mu_\ell \cos{b}$  & $\mu_b$  & K$_s$    & J & $\rm A_{K_{s}}$ & d   \\
        & (deg) & (deg) &  & (days) & (dex) & (arcmin) & ($\rm mas \ yr^{-1}$) & ($\rm mas \ yr^{-1}$) & (mag) & (mag) & (mag) & (kpc) \\
\hline
408\_103161 & 268.643884 & $-$23.993581 & RRL   & 0.904028  & $-1.791$ & 2.361 & -     & -        & 15.278 & 16.961 & 0.661 & 13.1 \\
608\_15410  & 268.638617 & $-$24.038442 & RRL   & 0.594803  & $-0.809$ & 2.711 & $-$3.434 & $-$2.944 & 14.809 & 16.422 & 0.629 & 8.4 \\
608\_18897  & 268.686531 & $-$23.976040 & RRL   & 0.554302  & $-1.473$ & 2.389 & $-$1.858  &  1.434    & 14.904 & 16.593 & 0.664 & 8.8 \\
608\_29612  & 268.680894 & $-$24.024242 & CEPII & 56.561086 & -        & 0.615 & $-$3.948 & $-$0.394 & 14.4220 & 16.3930 & 0.715 & 68.08 \\
608\_30628  & 268.687777 & $-$24.017228 & RRL   & 0.612895  & $-1.373$ & 0.658 & $-$1.483  &  2.977    & 15.900 & 17.448 & 0.598 & 14.5 \\
608\_45229  & 268.709713 & $-$24.037748 & RRL   & 0.532334  & $-1.183$ & 2.396 & $-$3.805  & $-$1.162 & 15.010 & 16.726 & 0.677 & 8.6 \\
608\_29840  & 268.682001 & $-$24.023419 & RRL   & 0.585514  & $-0.714$ & 0.599 & $-$9.493  & 0.863  & 15.460 & - & - & -\\
\hline
\end{tabular}
\label{table:tab3}
\end{table*}

\label{lastpage}
\end{document}